\documentclass[12pt]{article}
\usepackage{epsfig}
\usepackage{floatflt}
\usepackage{refmerge}

\title{
\Large
\bf \boldmath New measurement of the rare decay
$\phi \to \eta' \gamma$ with CMD-2
} 

\author{
R.R.~Akhmetshin\footnote{Budker 
Institute of Nuclear Physics, Novosibirsk, 630090, Russia}, 
E.V.~Anashkin\footnotemark[1], 
M.~Arpagaus\footnotemark[1], 
V.M.~Aulchenko\footnotemark[1]
\footnote{Novosibirsk State University, Novosibirsk, 630090, Russia}, \and 
V.Sh.~Banzarov\footnotemark[1],  
L.M.~Barkov\footnotemark[1] \footnotemark[2],
N.S.~Bashtovoy\footnotemark[1], 
A.E.~Bondar\footnotemark[1] \footnotemark[2], \and
D.V.~Bondarev\footnotemark[1], 
A.V.~Bragin\footnotemark[1],  
D.V.~Chernyak\footnotemark[1], 
S.I.~Eidelman\footnotemark[1] \footnotemark[2], \and 
G.V.~Fedotovitch\footnotemark[1] \footnotemark[2],   
N.I.~Gabyshev\footnotemark[1],
A.A.~Grebeniuk\footnotemark[1], 
D.N.~Grigoriev\footnotemark[1], \and
V.W.Hughes\footnote{Yale University, New Haven, CT 06511, USA},
P.M.~Ivanov\footnotemark[1], 
S.V.~Karpov\footnotemark[1],
V.F.~Kazanin\footnotemark[1] \footnotemark[2], \and    
B.I.~Khazin\footnotemark[1], 
I.A.~Koop\footnotemark[1], 
M.S.~Korostelev\footnotemark[1], 
P.P.~Krokovny\footnotemark[1] \footnotemark[2], \and
L.M.~Kurdadze\footnotemark[1] \footnotemark[2], 
A.S.~Kuzmin\footnotemark[1] \footnotemark[2],  
I.B.~Logashenko\footnotemark[1], 
P.A.~Lukin\footnotemark[1], \and
A.P.~Lysenko\footnotemark[1], 
K.Yu.~Mikhailov\footnotemark[1] \footnotemark[2],    
I.N.~Nesterenko\footnotemark[1],
V.S.~Okhapkin\footnotemark[1], \and
A.V.~Otboev\footnotemark[1], 
E.A.~Perevedentsev\footnotemark[1]\footnotemark[2], 
A.A.~Polunin\footnotemark[1], 
A.S.~Popov\footnotemark[1] \footnotemark[2], \and
T.A.~Purlatz\footnotemark[1] \footnotemark[2], 
N.I.~Root\footnotemark[1] \footnotemark[2], 
A.A.~Ruban\footnotemark[1], 
N.M.~Ryskulov\footnotemark[1], \and
A.G.~Shamov\footnotemark[1],  
Yu.M.~Shatunov\footnotemark[1], 
A.I.~Shekhtman\footnotemark[1], 
B.A.~Shwartz\footnotemark[1] \footnotemark[2], \and
A.L.~Sibidanov\footnotemark[1] \footnotemark[2],
V.A.~Sidorov\footnotemark[1],
A.N.~Skrinsky\footnotemark[1],      
V.P.~Smakhtin\footnotemark[1], \and
I.G.~Snopkov\footnotemark[1],
E.P.~Solodov\footnotemark[1] \footnotemark[2],
P.Yu.~Stepanov\footnotemark[1],      
A.I.~Sukhanov\footnotemark[1], \and
J.A.Thompson\footnote{University of Pittsburgh, Pittsburgh, PA 15260, USA}, 
V.M.~Titov\footnotemark[1], 
A.A.~Valishev\footnotemark[1], 
Yu.V.~Yudin\footnotemark[1], \and
S.G.~Zverev\footnotemark[1] 
}
\begin{document}
\maketitle


\newpage
\begin{abstract}
   A new measurement of the rare decay $\phi \rightarrow \eta' \gamma$ 
performed with the CMD-2 detector at Novosibirsk is described. Of the 
data sample corresponding to the integrated luminosity of 14.5 pb$^{-1}$, 
twenty one events have been selected in the mode $\eta'\to\pi^+\pi^-\eta$, 
$\eta\to\gamma\gamma$. The following branching ratio was obtained: 
\begin{center}
B($\phi \rightarrow \eta' \gamma) ~=~ (8.2^{+2.1}_{-1.9}\pm1.1) \cdot
10^{-5}$.
\end{center}

\end{abstract}
%

\section{Introduction}

\hspace*{\parindent}Radiative decays of vector mesons
have traditionally been a good laboratory for various tests
of the quark model and SU(3) symmetry \cite{don}. A recent
discovery of the $\phi \to \eta' \gamma$ decay by the
CMD-2 group \cite{fCMD} has been the last link in the otherwise
complete picture of radiative magnetic dipole transitions
between light vector and pseudoscalar mesons. This observation
was later confirmed by the SND group \cite{fSND}. Both
experiments suffered from a low number of observed events,
resulting in large uncertainties in the determined branching
ratio and making comparison to theory difficult.


   In this paper we report on the improved measurement
of the rate of the $\phi \to \eta' \gamma$ decay based upon
the total data sample accumulated with CMD-2 in the
$\phi$-meson energy range. It includes 3.1 pb$^{-1}$ of data
collected in 1992 -- 1996 in our first measurement which used
only photons observed in the CsI barrel calorimeter,
and about 11.4 pb$^{-1}$ collected in 1997 -- 1998.
In addition, this analysis uses photons detected in either the
CsI barrel or the BGO endcap calorimeters for both data samples 
providing better detection efficiency than before. 
 

The general purpose detector CMD-2 operating at the high luminosity
$e^+e^-$ collider VEPP-2M in Novosibirsk has been described in
detail elsewhere \cite{CMD285,cmd2gen}.
It consists of a drift chamber and proportional Z-chamber used for trigger, 
both inside a thin (0.4 $X_0$) superconducting solenoid with a field of 1 T.

The barrel calorimeter placed outside the solenoid consists of 892 CsI
crystals of $6\times 6\times 15$ cm$^3$ size and covers polar angles from
$46^\circ$ to $132^\circ$. The energy resolution for photons 
is about 9\% in the energy range from 50 to 600 MeV.

The end-cap calorimeter placed inside the solenoid consists of 680
BGO crystals of $2.5\times 2.5\times 15$ cm$^3$ size and covers 
forward-backward polar angles from 16$^\circ$ to 49$^\circ$ and 
from 131$^\circ$ to 164$^\circ$. The energy and angular resolution
are equal to $\sigma_E/E = 4.6\%/\sqrt{E(GeV)}$
and $\sigma_{\varphi,\theta} = 2\cdot10^{-2}/\sqrt{E(GeV)}$ radians
respectively.

The luminosity was determined from the
detected $e^+e^- \to e^+e^-$ events \cite{prep99}.



\section{\boldmath Decay kinematics and selection criteria}

\hspace*{\parindent}Since $\phi \rightarrow \eta' \gamma$ is a two-body 
decay and $\eta'$ is a narrow state, the momentum of the recoil photon 
is fixed and approximately equals 60 MeV.

To study this decay we searched for the decay chain 
$\eta' \to \pi^+\pi^-\eta$, $\eta \rightarrow \gamma \gamma$.
The photons are ordered by decreasing energy
($\omega_1 > \omega_2 > \omega_3$).
In these events the softest photon must be a
monochromatic recoil photon with the energy $\omega_3 \approx 60$ MeV
at the $\phi$ meson peak, while the energies of the harder ones range from
170 to 440 MeV.  
The invariant mass of the two harder photons 
$M_{12}~=~M_\eta~$.

The main source of background for this study  is the decay
mode $\phi \to \eta \gamma$ giving the same 
final state with two charged pions and three photons 
via the decay chain
$\eta \to \pi^+\pi^-\pi^0$, $\pi^0 \to \gamma \gamma$.
Here the hardest photon is
monochromatic with $\omega_1~=~363$ MeV and the invariant mass of two
others is $M_{23}~=~M_{\pi^0}$. This decay can be used as a
monitoring process and the branching ratio  
$B(\phi \rightarrow \eta' \gamma)$ will be calculated relative to 
$B(\phi \rightarrow \eta \gamma)$. Due to similar
kinematics and detection efficiency dependence on detector parameters
some systematic errors will cancel in such a ratio. 

Events with two tracks and three photons were selected using the following
criteria:

\begin{itemize}
\item{ One vertex is found in the event}
\item{ Two tracks with opposite charges are reconstructed from
this vertex and there are no other tracks}
\item{ The angles of both tracks with respect to the beam
are limited by $40^\circ< \theta <140^\circ$ to match the optimal drift
chamber coverage}
\item{ The number of photons detected in the CsI and BGO calorimeters
is three. The cluster in the calorimeter is
accepted as a photon when it does not match any charged track and its
energy is more than 30 MeV in the CsI calorimeter  or more than 40 MeV in
the BGO calorimeter.}
\item{ The distance from each track to the beam $R_{min}~<~0.2$ cm}
\item{ The distance from the vertex to the interaction point
along the beam direction $|Z_{vert}|~<~10$ cm}
\item{The space angle between the tracks $~\Delta \psi~<~
143^\circ$}
\item{The angle between the tracks in the R-$\varphi$ plane
$~\Delta \varphi~<~172^\circ$}
\item{ The total energy of the charged particles (assuming
that both particles are charged pions) $\varepsilon_{\pi^+\pi^-}<520$ MeV.}
\end{itemize}

The events thus selected were subject to the kinematical reconstruction
assuming energy-momentum conservation. 
Events with good quality of the reconstruction were selected by the following 
criteria:

\begin{itemize}
\item{$\chi^2/d.f.~<~3$}
\item{The ratio of the photon energy measured in the calorimeter
$\omega_{cal}$ to that from the constrained fit
$\omega$ is $\omega_{cal}/\omega~<~1.5$}
\item{$\omega_3 > 10$ MeV}
\end{itemize}


\begin{figure}[h]
\begin{center}
\includegraphics[width=0.8\textwidth]{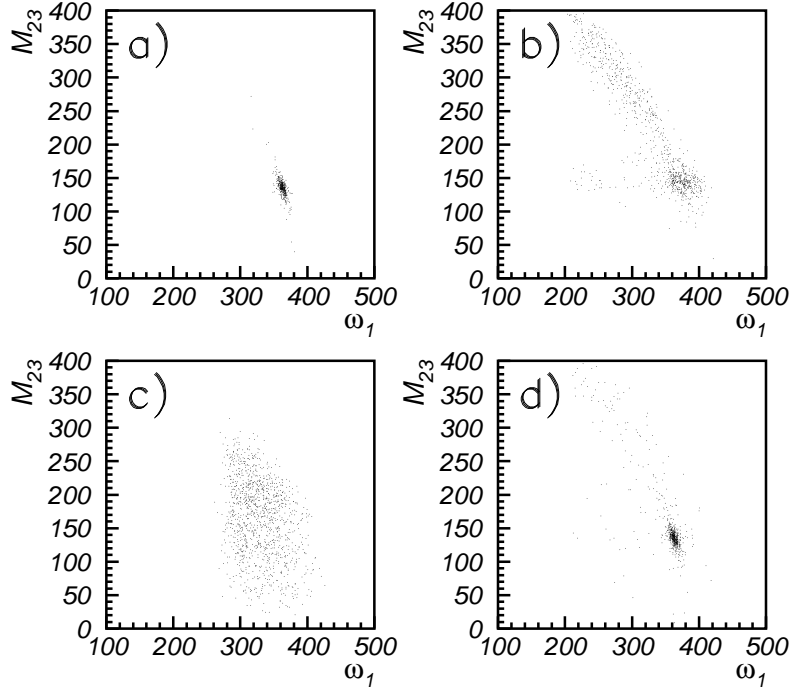}
\vspace{-1.cm}
\caption{Invariant mass of two soft photons $M_{23}$ vs hardest photon energy
$\omega_{1}$. { a)} --- simulation of $\phi \rightarrow \eta \gamma,
\eta \rightarrow \pi^+\pi^-\pi^0$;
{ b)} --- simulation of $e^+e^-\to\omega\pi^0\to\pi^+\pi^-\pi^0\pi^0$
at the $\phi$-meson energy;
{ c)} --- simulation of $\phi \rightarrow \eta' \gamma,
\eta' \rightarrow \pi^+\pi^-\eta, \eta\to\gamma\gamma$;
{ d)} --- experimental data.}
\label{fig:w1m23}
\end{center}
\end{figure}

\section{Analysis}

\hspace*{\parindent}Events surviving after all above criteria mostly 
come from the process $\phi\to\eta\gamma$, $\eta\to\pi^+\pi^-\pi^0$ and
$e^+e^-\to\omega\pi^0\to\pi^+\pi^-\pi^0\pi^0$, as illustrated 
by Fig.~\ref{fig:w1m23} 
showing the scatter plot of the invariant mass 
$M_{23}$ versus the hardest photon energy $\omega_1$. The data 
are shown in Fig.~\ref{fig:w1m23}{d}. 
The region around
$M_{23}~=~135$ MeV and $\omega_1~=~363$ MeV  is densely populated with
$\phi \rightarrow \eta \gamma, ~\eta \rightarrow \pi^+\pi^-\pi^0$ events.
Simulated events of this process are presented in 
Fig.~\ref{fig:w1m23}{a}. 
To determine the number of $\phi\to\eta\gamma$ events we
count the number of events inside the ellipse-like region:
$$\frac{(\omega_1~+~0.45 \cdot (M_{23}-m_{\pi^0})~-~\omega_
{\eta\gamma})^2}{22 \ MeV}+\frac{(M_{23}-m_{\pi^0})^2}
{60 \ MeV}~<~1.$$
For our data this number is $N_{\eta\gamma} = 7357$. Determination of
the number of $\eta\gamma$ events for simulation gives the detection
efficiency $\varepsilon_{\eta\gamma} = (15.5\pm0.3)\%$.

Figure~\ref{fig:w1m23}{b} 
presents the simulation
of $e^+e^-\to\omega\pi^0\to\pi^+\pi^-\pi^0\pi^0$, where 
a densely populated region is also observed at large values of $\omega_1$. 
Comparison of these distributions with that for the data 
(Fig.~\ref{fig:w1m23}{d}) 
confirms that the dominant contribution to selected events comes
from these two processes. 
The same distribution for the simulation of the process under study
is shown in Fig.~\ref{fig:w1m23}{c}.

\begin{figure}[h]
\begin{center}
\includegraphics[width=0.8\textwidth]{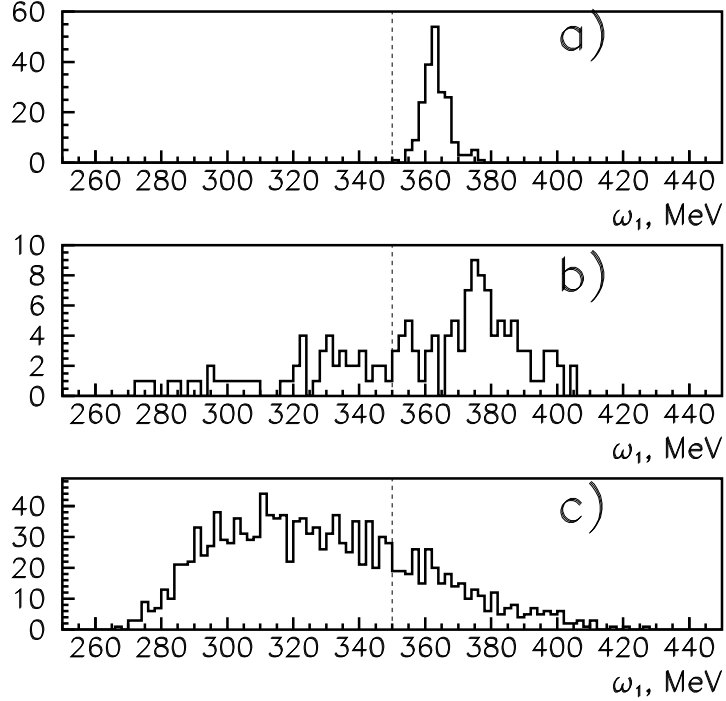}
\vspace{-1.cm}
\caption{The hardest photon energy $\omega_{1}$ for { a)} ---
simulation of $\phi \rightarrow \eta \gamma,
\eta \rightarrow \pi^+\pi^-\pi^0$;
{ b)} --- simulation of $e^+e^-\to\omega\pi^0\to\pi^+\pi^-\pi^0\pi^0$
at the $\phi$-meson energy;
{ c)} --- simulation of $\phi \rightarrow \eta' \gamma,
\eta' \rightarrow \pi^+\pi^-\eta, \eta\to\gamma\gamma$.}
\label{fig:wr1}
\end{center}
\end {figure} 
To search for the rare decay $\phi\to\eta'\gamma$ we need to suppress the 
events from $\phi\to\eta\gamma$ and $\omega\pi^0\to\pi^+\pi^-\pi^0\pi^0$.
To this end a cut on the energy of the hardest photon is applied: 
\begin{center}
 $\omega_1 < 350$  MeV.
\end{center}
The $\omega_1$ distributions for the
simulation of $\phi\to\eta'\gamma$ and background processes are shown in
Fig.~\ref{fig:wr1}.

\begin{figure}[h]
\begin{center}
\includegraphics[width=0.8\textwidth]{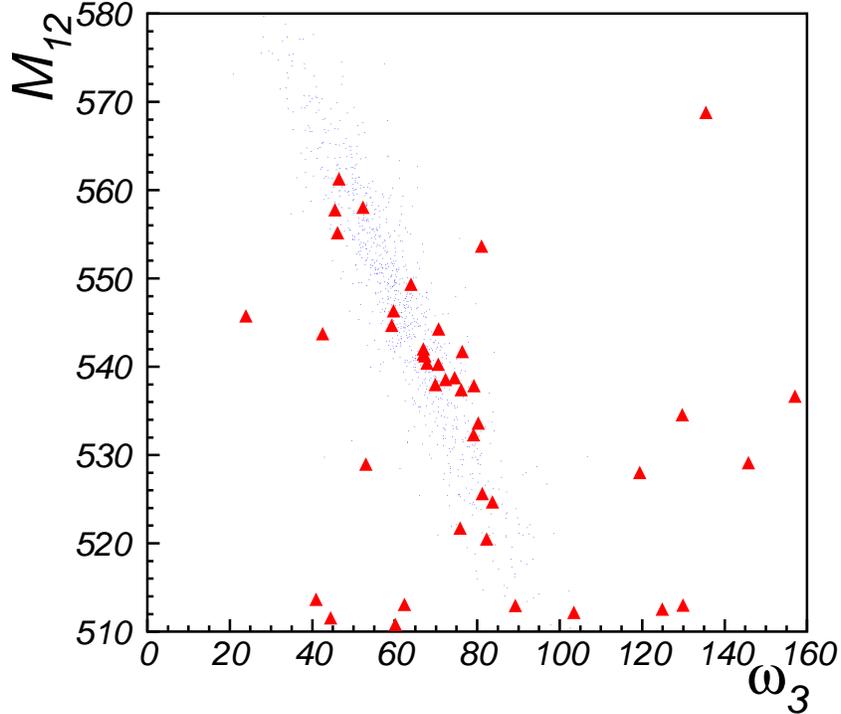}
\vspace{-1.cm}
\caption{Invariant mass of two hard photons $M_{12}$ vs softest photon
energy $\omega_{3}$. Points present the simulation of $\phi \rightarrow
\eta' \gamma, \eta'\to\pi^+\pi^-\eta, \eta\to\gamma\gamma$, triangles ---
data after all the selections.}
\vspace{.5cm}
\label{fig:finsum}
\end{center}
\end{figure}
Although this cut causes a decrease of efficiency for the $\phi 
\rightarrow \eta' \gamma$ decay (see Fig.~\ref{fig:wr1}c), 
the suppression of the background processes is rather good.

One more cut suppressing the background from the $\phi \to K_SK_L$ and
$\phi \to \pi^+\pi^-\pi^0$ decays is:
\begin{center}
 $\varepsilon_{\pi^+\pi^-}<420$ MeV.
\end{center}
After all the cuts the scatter plot of the invariant masses for two hardest 
photons $M_{12}$ versus the weakest photon energy
$\omega_3$ was studied. 
Figure~\ref{fig:finsum} 
presents the data (black triangles) 
together with simulation of $\phi \rightarrow \eta' \gamma$ (points). 
The simulation points show the region of the plot
which should be populated by the events of $\phi\to\eta'\gamma$
and experimental points are densely covering this region. The lower
part of the Figure contains obvious background events which can
be suppressed by imposing the additional cut $M_{12} > 515$ MeV. 

\begin{figure}[ht]
\begin{center}
\mbox{\psfig{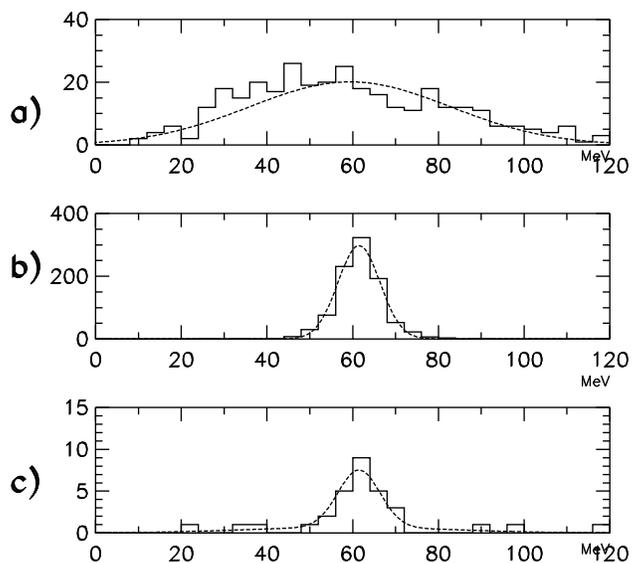}} 
\vspace{-0.5cm}
\caption{Distribution in $\omega_3+M_{12}-M_\eta$ together with the
fit function (dashed line): { a)} --- background from 
$\phi \rightarrow \eta \gamma, \eta \rightarrow \pi^+\pi^-\pi^0$ events; 
{    b)} --- simulation of 
$\phi \rightarrow \eta' \gamma, ~\eta' \rightarrow \pi^+\pi^-\eta, ~\eta
\rightarrow \gamma \gamma$; {    c)} --- data.}
\label{fig:etpr}
\end{center}
\end {figure}

To determine the number of events the one-dimensional distribution of
$\omega_3+M_{12}-M_\eta$ (projection of the plot in Fig.~\ref{fig:finsum} 
to the axis perpendicular to the correlation line) was studied. Such
projection is shown in Fig.~\ref{fig:etpr}{c} 
for the data. The same
projection for 10000 simulated events of $\phi \rightarrow \eta'
\gamma, ~\eta' \rightarrow \pi^+ \pi^- \eta, ~\eta \rightarrow \gamma
\gamma$ is shown in Fig.~\ref{fig:etpr}{b}, 
and the fit of this
distribution fixes the signal shape and gives the
detection efficiency $~\varepsilon_{\eta' \gamma}~=~(9.1 \pm 0.3)\%$.
The background distribution in this parameter determined from the 
data before applying the last two cuts 
($\omega_1 < 350$ MeV and $\varepsilon_{\pi^+\pi^-}<420$ MeV) 
 is shown in Fig.~\ref{fig:etpr}{a}.
The fit of
this distribution fixes the background shape. 
Finally, the data were fit using the background shape fixed from
Fig.~\ref{fig:etpr}{a} 
together with that of the signal from simulation in 
Fig.~\ref{fig:etpr}{b}.

The result of the fit is $N_{\eta'\gamma}~=~21.0^{+5.5}_{-4.9}$.

Using the number of events from the fit, one can calculate the 
relative branching ratio:

\vspace{.6cm}

$\frac{B(\phi \rightarrow \eta' \gamma)}{B(\phi \rightarrow \eta \gamma)}
~=~\frac{N_{\eta' \gamma}}{N_{\eta \gamma}} \cdot \frac{B(\eta \rightarrow 
\pi^+ \pi^- \pi^0)}{B(\eta' \rightarrow \pi^+ \pi^- \eta)} \cdot \frac
{B(\pi^0 \rightarrow \gamma \gamma)}{B(\eta \rightarrow \gamma \gamma)}
\cdot \frac{\varepsilon_{\eta \gamma}}{\varepsilon_{\eta' \gamma}}~=~
(6.5^{+1.7}_{-1.5})\cdot 10^{-3}$
\\

\vspace{.2cm}

\noindent
where the values of the branching ratios of $\eta', \eta$ and $\pi^0$
were taken from \cite{pdg}.

A separate analysis of the normalizing decay $\phi \to \eta \gamma$ 
has recently been published \cite{etag}, with a branching ratio of 
$(1.18 \pm 0.03 \pm 0.06)\%$
consistent with previous measurements \cite{pdg} and thus giving
confidence in the analysis presented here. 

In the above calculation of the relative
branching ratio common systematic errors such as 
luminosity determination cancel exactly, while others such as 
detector inefficiency and evaluation of radiative corrections 
cancel approximately.


Finally, using the value of $B(\phi \to \eta \gamma) = (1.26 \pm 0.06)\%$
from \cite{pdg}, one obtains:
\begin{center}
{$B(\phi \rightarrow \eta'
\gamma)~=~(8.2^{+2.1}_{-1.9}\pm1.1)\cdot 10^{-5}$.}
\end{center}

The last error is our estimate of the systematic uncertainty. The sources of 
systematic errors are the following:
\begin{itemize}
\item
Uncertainties in the ratio $\frac {\varepsilon_{\eta \gamma}}
{\varepsilon_{\eta' \gamma}}$ caused by different energy spectra
for final photons and pions - 10\%;
\item Uncertainties in the branching ratios 
$B(\eta \rightarrow \pi^+ \pi^- \pi^0)$, 
$B(\eta' \rightarrow \pi^+ \pi^- \eta)$,
$B(\eta \to \gamma \gamma)$ and $B(\phi \to \eta \gamma)$ - 6.3\%;
\item Determination of the background shape -- 5\%;
\item Different resonance shape caused by different energy
dependence of the phase space  - 2\%;
\end{itemize}

The total systematic error obtained by adding separate contributions
quadratically is 13\%.

\section{Discussion}

\hspace*{\parindent}The results of our analysis have higher 
statistical significance than before
since they are based on a data sample of 21 selected events 
compared to 6 events in our previous work \cite{fCMD} and 5 events observed
by SND \cite{fSND}. 
The obtained value of the branching ratio $B(\phi \to \eta' \gamma)$ 
\begin{center}
$(8.2^{+2.1}_{-1.9} \pm 1.1) \cdot 10^{-5}$. \\
\end{center}
agrees with our previous result based on part of the
whole data sample \cite{fCMD} 
\begin{center}
$(12.0^{+7.0}_{-5.0}\pm1.8)\cdot 10^{-5}$ \\
\end{center}
as well as with the result of the SND group \cite{fSND} 
\begin{center}
$(6.7^{+3.4}_{-2.9}\pm1.0)\cdot 10^{-5}$ \\
\end{center}
and is more precise. Within experimental accuracy it is also
consistent with the preliminary result of CMD-2 based on other
decay modes of the $\eta$ meson 
($\eta \to \pi^+\pi^-\pi^0, \pi^+\pi^-\gamma$) with four charged pions
and two or more photons in the final state \cite{prep99}: 
\begin{center}
$(5.8 \pm1.8 \pm1.5)\cdot 10^{-5}$. 
\end{center}

Analysis of the available data sample of the produced $\phi$ mesons by
both CMD-2 and SND and full use of other decay modes of the $\eta'$ and
$\eta$ mesons will further improve the statistical error. Much larger
increase can be expected from the DA$\Phi$NE $\phi$-factory where
one plans to accumulate the number of $\phi$ mesons by at least two 
orders of magnitude higher than ours. 

Let us briefly discuss theoretical predictions for the decay under study. 
Usual methods of the description of radiative decays are
based on the nonrelativistic quark model \cite{don}. Various
ways of incorporating effects of SU(3) breaking have been suggested
leading to the values of $B(\phi \rightarrow \eta' \gamma)$
in the range $(5-20) \cdot 10^{-5}$ 
\cite{geffen,ohshima,ivan,bramon,ben95,ben96,ball,hashimoto,ko,feld,escri,ben99}.  

The value of the branching ratio studied in our work is also of interest
for the problem of $\eta-\eta'$ mixing which has been a subject 
of intense investigation for a long time 
\cite{nsvz,rosner,gk,af,aek,bes,feld99,bag}.
It is sensitive to the structure of the $\eta'$ wave function or,
in other words, to the contribution of various $q\bar{q}$ states 
as well as the possible admixture of glue in it \cite{desh,rosner,kou}.  
According to \cite{desh}, a branching ratio 
$B(\phi \rightarrow \eta' \gamma) < 2 \cdot 10^{-5}$ would indicate
a substantial glue component in the $\eta'$, while the expected branching 
ratio is less than $3 \cdot 10^{-6}$ for a pure gluonium. Even smaller 
values were obtained in \cite{ben95} assuming a specific model 
of QCD violation. 

The revival of interest to the
problem of the $\eta'$ structure and possible contents of glue in it
(see \cite{kou} and references therein)
was partially due to two recent observations by CLEO involving
the $\eta'$ meson: in \cite{gron} it was shown that the transition
form factor of the $\eta'$ studied in the two photon processes
strongly differs from those for the $\pi^0$ and $\eta$ mesons and
in \cite{betap} the unexpectedly high magnitude of the rate
of $B \to \eta' K$ was observed.   
However, in a recent paper \cite{ben99} 
it is claimed that it is impossible to disentangle the effects
of the nonet symmetry breaking and those of glue inside the $\eta'$.

Most of the models mentioned above are able to describe the data 
reasonably well in terms of some number of free parameters
which, unfortunately, can not be determined from first
principles. An attempt to overcome this drawback was made in \cite{zhu} 
where radiative decays of
light vector mesons are considered in the approach based on QCD sum 
rules \cite{qcd} and the value $15 \cdot 10^{-5}$ is obtained
for the branching ratio of $\phi \to \eta' \gamma$ decay. 

One can summarize that the variety of theoretical approaches
to the problem of the description of the $\phi$ meson radiative decay 
to $\eta' \gamma$ is rather broad and more theoretical insight 
into the problem is needed.
  
\section{Conclusions}

\hspace*{\parindent}Using an almost five times bigger data sample 
than in the first measurement the CMD-2 group 
confirmed the observation of the rare radiative decay
$\phi \rightarrow \eta'\gamma$. The measured branching
ratio is: 
\begin{center}
{$B(\phi \rightarrow \eta'
\gamma)~=~(8.2^{+2.1}_{-1.9}\pm1.1)\cdot 10^{-5}$.} \\
\end{center}

Its value is consistent with most of the theoretical 
predictions based on the quark model and assuming 
a standard quark structure of the $\eta'$. It
rules out exotic models suggesting a high glue admixture
\cite{desh} or strong QCD violation \cite{ben95}. 

Further progress in this field can be expected after the 
dramatic increase of the number of produced $\phi$ mesons 
expected at the DA$\Phi$NE $\phi$-factory and refinement
of theoretical models of radiative decays.  

\section{Acknowledgements}

\hspace*{\parindent}The authors are grateful to M.Benayoun and 
V.N.Ivanchenko for useful discussions.

\end{document}